\documentstyle[12pt,aasms4]{article}

\def\gappr{\mathrel{\vcenter{\offinterlineskip \hbox{$>$}
    \kern 0.3ex \hbox{$\sim$}}}}
\def\lappr{\mathrel{\vcenter{\offinterlineskip \hbox{$<$}
    \kern 0.3ex \hbox{$\sim$}}}}

\received{}
\revised{}
\singlespace

\begin{document}

\title{The role of thermal instabilities on the theoretical \\
mass function of MACHOs}

\author{E.M.de Gouveia Dal Pino , J.E.Horvath and
J.A.de Freitas Pacheco$^{1}$}

\affil{Instituto Astron\^{o}mico e Geof\'{\i}sico, Universidade
de S\~{a}o Paulo, \\
 Av. Mguel Stefano 4200,
(04301-904) S\~{a}o Paulo - SP, BRAZIL \\
E-mail: dalpino@astro1.iagusp.usp.br}

\singlespace
\affil{$^{1}$ also at Observatoire Cote D'Azur,
BP 229 Nice Cedex 4 - 06304 FRANCE}

\begin{abstract}

We analyze the primordial protogalactic gas to investigate the role of
thermal instabilities in the formation of condensations that could
contribute to the MACHO mass function. We find that the previrialized
protogalaxy could have undergone a thermally unstable regime which
defines a peak in mass condensations
starting around $0.1 \, M_{\odot}$ after being reprocessed by the environment
which may be identified as giving origin to the MACHOs. Effects of
magnetic fields on this instability-driven MACHOs scenario are also addressed.
\end{abstract}


\clearpage

\section{Introduction}

Two years after the first reports of candidate microlensing events
(Alcock et al. 1993, Aubourg et al. 1993, Udalski et al. 1993) the field
is rapidly growing and bocoming a major tool to investigate the morphology
and history of our galaxy (see Zylberajch 1994 and Paczy\'nski 1995 for
reviews). Particularly intriguing is the fact that a massive halo composed of
MACHOs seems {\it not} to be favored by the observations (Alcock et al. 1995,
Gates et al. 1995). The nature of dark objects so far responsibles for the
events towards the LMC needs also to be addressed. It is not clear as yet
if they belong to a disk of our own galaxy, to the galactic halo
itself, or even to the LMC (Sahu 1994). We shall adopt in this work the
hypothesis that MACHOs are indeed located in the halo and attempt to
outline a scenario that could have leaded to their formation.

A number of works have stressed the possible relevance of {\it thermal
instabilities} (TI) for globular cluster formation during the collapse
of the protogalactic cloud (e.g., Fall \& Rees 1985,
 Vietri \& Pesce 1995 and references therein).
While these scenarios address an epoch in which the gas has already been
shock-heated to its virial temperature $T \, \sim \, 10^{6} \, K$,
we will be mainly interested in a previous, optically thin stage of the
cloud at $T \, \sim \, few \, \times \,10^{4} \, K$.
We also note some previous works dealing with the protogalaxy which have
addressed the question of Pop III star formation (e.g. Silk 1983).
The motivation for a reasessment of such scenarios is stronger than ever,
particularly now that an initial spectrum can be calculated and its
reprocessing characterized observationally by the growing body of data.
To stress this point, let us consider, as in those previous
works, a quasi-static, isothermal protogalactic cloud with an estimated
virial temperature $T \, \sim \, \mu v_{r}^{2}/2 k_{B}$, or about $10^{6} K$.
It is known that at earlier stages the faster cooling at the
center would have destroyed the dynamical
equilibrium and induced the collapse of
the cloud, mainly because the equilibrium requires the opacity of the gas
to be large enough in order to delay radiative losses. On the other hand,
an optically thin gas will cause the cloud to cool more quickly, so
that at least for timescales of the order of the free-fall time
$t_{ff} \, \sim \, (G \rho)^{-1/2}$ the protogalactic gas will remain at a
temperature much smaller than the virial one. There is also an observational
motivation to consider a previous epoch of the galaxy; namely that a
substantial fraction of the halo mass is in the form of field stars, not
in globular clusters at a ratio $1/100$. Furthermore, the most metal-poor
clusters have $[Fe/H] \, \simeq \, -2.5$, while some field stars have been
detected with $[Fe/H] \, \simeq \, -4$, suggesting that they were formed
$before$ globular clusters.

\section {Thermal instability}

Motivated by the above discussion we consider  a previrialized collapsing
cloud with mass $M \, = \, 10^{12} \, M_{\odot}$, radius
$r \, \gappr \, 50 \, kpc$
(corresponding to an average density
$\rho \, \simeq \, 10^{-25} \, g \, cm^{-3}$) and
$T \, = $ few $\times \, 10^{4} \, K$. Such a cloud would have started to
collapse at $z \, \leq \, 10$ when its mean density was about $5$ times the
background density (Rees \& Ostriker 1977).

At temperatures below $T \; \sim \; 3 \times 10^{4} K$
the primordial gas of the cloud (which has a metal abundance $Z \, \sim \, 0$)
cooling through radiative thermal processes
(collisional excitation and recombination) is stable to condensation formation
by isobaric thermal instability (e.g. Field 1965, Corbelli \& Ferrara 1995).
We assume
that the onset for the thermal instability occurs at
$T \, \gappr 3 \times 10^{4} K$.
We also assume a perfect gas equation of state
$P = R \, \rho \, T/\mu$ , where $\mu \, = <m>/m_{H}$ and $\rho,T$ are the
equilibrium density and temperature of the cloud.
A perturbed gas element is subject to an isobaric
thermal instability growing to the first order at a rate
(Field 1965, see also Begelman 1990, Gouveia Dal Pino \& Opher 1989a,b, 1990,
1991, 1993)
\begin{equation}
 n  = \biggl[
{{\partial {\cal L} \over {\partial T}} \vert_{\rho}} \, - \,
{\rho \over {T}} {\partial {\cal L} \over {\partial \rho}}
\vert_{T} \, + \,
{k^{2} K \over {\rho}} \biggr]
{(1- \gamma) \mu \over {\gamma R}}
\end{equation}

\noindent
where $\mu$ is the mean molecular weight ($= \, 0.59$ for an ionized
gas of primordial
composition), $\gamma = 5/3$,
$K = 0.77 \times 10^{-6} \, T^{5/2}$ is the thermal
conductivity term  and ${\cal L} \, = \, L(\rho,T) \, -\, G(\rho,T)$, i.e.,
cooling minus heating, is the net
energy loss per unit time and mass, which vanishes at equilibrium. Note
that, strictly speaking, eq.(1)
holds for the short wavelenghts of the perturbation.
A very important (and uncertain) point is the nature and dependence of the
heating function $G(\rho,T)$. The heating function may include the UV flux
of an AGN phase (likely to have been present in the early galaxy because
of the observational evidence collected from the bulge) ; turbulence
(the Reynolds number
$Re \, = \, 2 \, r_{c} \, v /\eta $ is known to be $ \gg \, 1$ for
the primordial gas, $v$ being the gas velocity and $\eta$ the shear viscosity);
  and MHD waves
among the most important sources.
Given the uncertainties on the nature of the dominant
heating mechanisms in the
protogalactic gas, we assume as in previous works
(e.g. Field 1965, Eilek \& Caroff 1979, Gouveia Dal Pino \& Opher 1989a,1989b,
1990, 1991, 1993) that the heating function $G (\rho,T)$ is due to an
externally-generated source and its dissipation is locally independent of
the density and temperature, a condition which is known to hold for several
of the above mentioned sources.
Thus, the derivatives of ${\cal L}$ in
eq.(1) will depend only on the cooling function $L$.

For a strongly ionized collisional $H \, + \, He$ plasma of primordial
composition
( $X \, \simeq \, 0.76$, $Y \, \simeq \, 0.24$ and $Z \, \simeq \, 0$) at $
2.5 \times 10^{4} K \, \lappr T \lappr
10^{6} K$ the net cooling is
\begin{equation}
L \, \simeq \,  \Lambda (T) \, n_{H}^{2} / \rho \; \, erg \,
g^{-1} s^{-1} ,
\end{equation}

\noindent
where $n_{H}$ is the hydrogen number density and the cooling
function $\Lambda(T)$, evaluated by, e.g.,
 Fall \& Rees
(1985),  can be fitted by
\begin{equation}
log \Lambda (T) \, = \, -22.9 \, + \, 2.45 \, \phi(T) \, exp(-\phi(T))^{2} \, +
\, \psi(T) \, exp(-\psi(T))^{2} \, + \, \theta(T) \, exp(-\theta(T))^{2}
\end{equation}

\noindent
with $ \phi(T) \, = \, 1.1 \, log T \, - \, 4$ ,
$\psi(T) \, = \, 3.7 \, log T \, - \, 17.6$ and
$\theta(T) \, = \, 1.4 \, log T \, - \, 9.2$. This rather complicated fit
can be replaced by the much simpler expression
$log \Lambda(T) \, = \, -24.45 \, + 1.04 \, log T \, - \, 0.11 (log T)^{2}$
in the restricted region of interest of this work
$2.5 \times 10^{4} K \, \lappr \, T \, \lappr \, 10^{5} K$
with sufficient accuracy. Furthermore, since
${\left\vert {T \partial {\cal L}/ \partial T} \right\vert} \, \ll \,
{\left\vert {\rho \partial {\cal L}/ \partial \rho} \right\vert}$ (see eq.1),
we can safely neglect $\partial {\cal L}/ \partial T$ in our analysis and
adopt a constant $\Lambda(T) \, \simeq \, 9.6 \times 10^{-23} \, erg \, cm^{3}
\, s^{-1}$ in eq.(2); this extremely simple form introducing errors of
at most $10 \, \%$ in the calculated quantities.

An isobaric thermal instability grows only if the pressure in the perturbed
region is equalized in a time-scale small compared to the cooling time,
defined as $t_{c} \,
\equiv \, (3 \, \rho^2 \, k_{B} \, T)(2 \, \mu \, m_{H} \, L)^{-1}$.
This implies an upper limit for the wavelength of the unstable
region (e.g. Field 1965)
\begin{equation}
\lambda \, \leq \, \lambda_{max} \simeq \ 2 \, \pi \, v_{s} \,
t_{c} \, = 1.1 \, \times \, 10^{20} \, (\rho_{-25})^{-1} \;
(T_{4.7})^{3/2} \, cm.
\end{equation}
where $v_{s} \, = \, (\gamma \, k_{B} \, T / \mu \, m_{H})^{1/2}$ is the
speed of sound,
$\rho_{-25} =
\rho / (10^{-25} \,
g \, cm^{-3})$, and
$T_{4.7} = T / (5 \, \times \, 10^{4} K)$.

On the other hand, it is easy to show that,
due to the thermal conduction ability to supress
the growth of a thermally unstable region,
 a critical wavenumber $k_{max} = \, 2 \, \pi / \lambda_{min}$
exists over which
$n \, = \, 0$ and it is given by
$k_{max} = \, {\bigl[ k_{K} \, (k_{\rho} - k_{T}) \bigr]}^{1/2}$ ;
where
$k_{\rho} = {(\gamma - 1) \, \mu \, \rho
{\bigl( {
\partial {\cal L} \over {\partial \rho}}\bigr)_T} \, \over{  R
 \, v_{s} \, T }}$,
$k_{K} = {{R \, v_{s} \, \rho} \over{{\mu \, (\gamma -1) \, K}}}$, and
$k_{T} = {(\gamma - 1) \, \mu \,
{\bigl( {
\partial {\cal L}
\over {\partial T}}\bigr)_\rho }  \over { R \,
 v_{s}}}$.
After an elementary manipulation with these expressions we find
$\lambda_{min} \, = \, 2 \, \pi / k_{max}$ which is given by
\begin{equation}
\lambda_{min} \, \simeq \,  2.1 \times 10^{18}
(\rho_{-25})^{-1} \,
(T_{4.7})^{7/4}  \, cm .
\end{equation}
Since $n$ is small for small $k$ (large $\lambda$) and vanishes for
$k \, > \, k_{max}$, a $k$ exists for which $n$ reaches a maximum. This
value is
$k(n_{max}) \, \simeq \, {\bigl[ k_{\rho}^{2} (k_{\rho} \, - \, k_{T})
k_{K} \bigr]}^{1/4}$, or in terms of $\lambda (n_{max})$
\begin{equation}
\lambda (n_{max}) \, \simeq \, 1.5 \times 10^{19}
(\rho_{-25})^{-1} \,
(T_{4.7})^{13/8} \, cm .
\end{equation}
The most rapidly growing condensations are those at
$\lambda (n_{max})$. They define a typical mass
$M_{blob} \, \sim \, \rho_{b}
\, \lambda^{3}(n_{max})$ or
\begin{equation}
M_{blob} \, \sim \, 1.8 \times 10^{-1} \,
{\bigl( 1 \, + \, f_{b} e \bigr)}
(\rho_{-25})^{-2} \,
(T_{4.7})
^{39/8} \, M_{\odot} ,
\end{equation}
\noindent
where $f_{b} \, < \, 1$ is the initial density perturbation relative to the
ambient density $\rho$.
This is the most probable mass of the condensations, which is bracketed
between the limiting values
\begin{equation}
M_{blob} (\lambda_{min}) \, \sim \, 4.4 \times 10^{-4} \,
{\bigl( 1 \, + \, f_{b} e \bigr)}
(\rho_{-25})^{-2} \,
(T_{4.7})
^{21/4} \, M_{\odot} ,
\end{equation}
\noindent
and
\begin{equation}
M_{blob} (\lambda_{max}) \, \sim \, 73 \,
{\big( 1 \, + \, f_{b} e \bigr)}
(\rho_{-25})^{-2} \,
(T_{4.7})
^{9/2} \, M_{\odot} ,
\end{equation}
\noindent
which define the condensation mass spectrum at early times.
\section{Primordial magnetic field effects}

If a small primordial field is present in the gas {\bf B} $\not= 0$
(which seems unavoidable in the pregalactic medium, e.g. Rees 1994)
in the direction $\parallel$ to {\bf B}, the condensation is not affected
by the presence of the magnetic field and its growth is given by eq. (1).
However, in the direction $\perp$ to {\bf B}, the thermal conduction
$K$ is greatly reduced (because of the $e^{-}$ spiraling
between collisions) and the initial collapse of the condensation is more
efficient (e.g. Field 1965 ;
Gouveia Dal Pino \& Opher 1989a, 1989b, 1990, 1991, 1993).
We can therefore determine two scales in this direction $\lambda_{\perp min}$
and  $\lambda_{\perp max}$ for the instability growth with analogous meanings
to those of the former section. Defining an effective pressure given by
the sum of the gas pressure and the magnetic pressure we have (e.g. Field
1965)
\begin{equation}
\lambda_{\perp min} \, \sim 4.1 \times 10^{17} \,
(B_{-12})^{-1} \,
(T_{4.7})^{1/4} \,
{\bigl( 1 \, + \, {\gamma v_{A}^{2} \over {v_{s}^{2}}} \bigr)}^{1/2} \,
cm ,
\end{equation}
\noindent
where $v_{A} \, = \, B/(4 \, \pi \, \rho)^{1/2}$ is the Alfven speed,
$B_{-12} =  B / (10^{-12} \, G)$,
and (e.g. Gouveia Dal Pino \& Opher 1990)
\begin{equation}
\lambda_{\perp max} \, \sim \, 1.1 \times 10^{20} \,
(\rho_{-25})^{-1} \,
(T_{4.7})^{3/2} \,
{\bigl( 1 \, + \, {v_{A}^{2} \over {v_{s}^{2}}} \bigr)}^{1/2} \, cm .
\end{equation}
Note that for small {\bf B} the ratio
$v_{A}^{2}/v_{s}^{2} \, \ll \, 1$ for typical values of protogalactic
$\rho$ and $T$.
The masses that bound the spectrum ($\lambda_{\parallel} \, = \,
\lambda (n_{max})$) can be estimated as
$M_{blob} (\lambda_{\perp min}) \, \simeq \, \rho_{b} \; \pi \;
\lambda_{\perp min}^{2} \;
\lambda_{\parallel}$, which results in
\begin{equation}
 M_{blob} (\lambda_{\perp min}) \, = \, 4.1 \times 10^{-4} \,
{\bigl( 1 \, + \, f_{b} e \bigr)}
(T_{4.7})
^{17/8} \,
(B_{-12})^{-2} \,
{\bigl( 1 \, + \, {\gamma v_{A}^{2} \over {v_{s}^{2}}}
\bigr)} \;
M_{\odot} ,
\end{equation}
\noindent
and $M_{blob} (\lambda_{\perp max}) \, \simeq \, \rho_{b} \; \pi \;
\lambda_{\perp max}^{2} \; \lambda_{\parallel}$, yielding
\begin{equation}
M_{blob} (\lambda_{\perp max}) \, = \, 31 \,
{\bigl( 1 \, + \, f_{b} e \bigr)}
(\rho_{-25})^{-2} \,
(T_{4.7})^{37/8} \,
{\bigl( 1 \, + \, {v_{A}^{2} \over {v_{s}^2}} \bigr)} \, M_{\odot} .
\end{equation}
\section{Evolution of the blobs}

{}From eq.(1) we can evaluate the  growth time $\tau_{g}$
of the condensations in the absence of {\bf B}:
\begin{equation}
\tau_{g} = {\biggl[ 1 \times 10^{-13}
(\rho_{-25}) \,
(T_{4.7})
^{-1}
- 1.2 \times 10^{22} (T_{4.7})
^{5/2}
(\rho_{-25})^{-1} k^{2}
\biggr]}^{-1} \, s
\end{equation}
\noindent
for $k \, \simeq \, k (n_{max})$ (eq.(6)) and $T,\rho$ close to our
scaling values, the first term dominates and gives $\tau_{g} \, \simeq \,
2.9 \, \times \, 10^{5} \, yr$. This initial growth
time-scale is much smaller than the
free-fall time of the condensed blobs
$\tau_{ff,b} \, \sim \, (G \, \rho_{b})^{-1/2}$ and indicates that
initially the condensations will not collapse gravitationally, but rather
contract and cool because of the  thermal instability itself.
(In the presence of {\bf B}
the time is given by $\tau_{g,B} \, \simeq \, 9 \, \times 10^{12}
T_{4.7} \,
(\rho_{-25})^{-1}
{\bigl( {1 \, + \, {v_{A}^{2} \over {v_{s}^{2}}}} \bigr)} \, s $, which for
$v_{A}^{2}/ v_{s}^{2} \, \ll \, 1$ is also of the order of $\tau_{g}$.)
In fact, the condensed blob will become gravitationally unstable
if its mass exceeds the
Jeans mass $M_{J} \, = \,
1.2 \, {(k_{B} \, T_b/ \mu m_{H})}^{2} \, G^{-3/2} \, P^{-1/2}$
(where $T_{b}$ is the initial
temperature of the blob and $P$ is the pressure of
the unperturbed medium), which is much larger
than $M_{blob}$ for temperatures $\, \simeq \,  10^4$ K.

A blob in pressure equilibrium with the ambient gas will have a density
$\rho_{b} \, = \, (T \, \mu_{b} / T_{b} \, \mu) \, \rho$ (where
$\mu_{b} \, \sim \, 1.22$ for a weakly ionized gas inside the blob).
Once the temperature of the blob dropped below $10^{4} K$, the
cooling will be dominated by the collisional excitation of the rotational
and vibrational transitions of the $H_{2}$ molecule (the heavy elements also
contribute, but for very small abundances $Z/Z_{\odot}$ they do not
dominate the
cooling,
see, e.g.,  Fall \& Rees 1985). Thus, just after its formation the
 blob will cool in
a time $\tau_{c,i} \, \sim \, (3 \, \rho_{b} \, k_{B} \, T_{b})
(2 \, \mu_{b} \,
m_{H} \, L_{H_{2}})^{-1}$ ; with $L_{H_{2}} \, \simeq \, 10^{-22} \,
n_{b}(H_{2}) \, erg \, cm^{-3} \, s^{-1}$ for an initial  hydrogen
number density
in the blob
$n_{b}(H) \, \sim \, 1 \, cm^{-3}$ (e.g. Lepp \& Shull 1983).
The number density
of molecular hydrogen $n_{b}(H_{2})$ can be evaluated from statistical
equilibrium taking into account the relevant reaction rates and the effect of
UV dissociation due to the hotter ambient gas; we find
$n_{b}(H_{2}) \, \simeq \, 3 \times 10^{-2} \, x \, n_{b}(H)$ at
$T_{b} \, \sim 9000 \, K$, where $x \, \equiv \, n_b(H^+)/n_{b}(H) \, < 1$
is the fractional ionization
(see, e.g., Fall \& Rees, eq.(18) and references therein).
In this estimate we have assumed a photoionization rate
$k_{UV} \, \simeq \, 2.4 \times 10^{-13} \, (r/kpc)^{-1} \, s^{-1}$,
appropriate for
our physical conditions. This cooling time
$\tau_{c,i} \, \simeq \, 2.0 \times 10^4 \, x^{-1} \, (T_b/9000 K) yr $
is now
larger than the condensation formation
time $\tau_{g}$ (eq. 14) for $x  \lappr 10^{-2}$
(e.g. Dalgarno \& McCray 1972, Fall \& Rees 1985), but still much smaller
than the protogalactic free-fall time
$t_{ff} \, \sim \, 4 \times 10^{8} \, (\rho / 10^{-25} \, g
\, cm^{-3})^{-1/2} yr$, so that the blobs will continue to cool within the
protogalactic infall gas.
A cold blob moving relative to the ambient gas may be subject
to the Kelvin-Helmholtz (K-H) instability
at the interface between blob and ambient gas with a characteristic
growth time $\tau_{K-H} \, \simeq \, \lambda_{b} (\rho_{b}/\rho)^{1/2}/ u$,
where $\lambda_{b}$ is the radius of the blob and $u$ is the relative
velocity (Murray et al. 1993). Buoyancy effects may accelerate some blobs
to speeds comparable to the  sound speed, for which
$\tau_{K-H} \, \sim \, \lambda_{b} / (\gamma \, k_{B} \, T_{b} /
\mu_{b} m_{H})^{1/2}$. Imposing the "survival"
condition  for a blob against K-H instability
$\tau_{c,i} < \, \tau_{K-H}$ gives
a lower limit to the size of the blobs
$\lambda_{b} > 6.3 \times 10^{17} \, x^{-1} \,
(T_{b}/9000 K)^{3/2} \, cm $.
 Thus, blobs moving with relative speed close
to the sound speed are expected to break up since they do not satisfy this
condition for $x \lappr 10^{-2}$ (eqs. 6, 10, 11). On the other hand,
those "moving with" the collapsing protocloud will be
stable to the K-H instability. The presence  of magnetic
fields will further inhibit the K-H instability effects, although the surviving
fraction of blobs {\it must not} be
expected to be large since in that case the observed
MACHO fraction would also be large which is not the case
(Alcock et al. 1995, Gates et al.1995). A detailed calculation of the K-H
effects  is out of the scope of this work and will be discussed elsewhere.

Finally, the blobs must survive to a much "dangerous" later epoch in which the
protocloud has been heated to $T \, \sim \, 10^{6} K$ (for some radius)
and $\rho \, \sim \, 10^{-24} \, g \, cm^{-3}$ (e.g. Fall \& Rees 1985).
The blobs which should have by now cooled to
$T_{b} \, \lappr \, 1000 \, K$ since
$\tau_{c,f} \, \simeq \, 13 \, n_{b}(H) / n_{b}(H_{2}) \, yr$
(where we have assumed
$L_{H_{2}} \, \simeq \, 5 \times 10^{-22} \, n_b(H_{2})
 \,  erg \, cm^{-3} \, s^{-1}$ for $n_{b}(H) \, \sim 10^{3} \, cm^{-3}$;
see Lepp \& Shull 1983) will be heated by X-rays from the hot ambient
protocloud. The effective cross section for photoionization of the blob gas
due to $\sim \, 0.1 \, KeV$ photons is $\sigma \, \simeq \, 1.8 \times
10^{-20} \, (h \nu / 150 \, eV)^{-3} \, cm^{2}$. Since the radiation is
mainly thermal bremsstrahlung with an intensity
$ \propto \, exp(- h \nu / k_{B} T)$, about $90 \, \%$ of the heat will
be deposited in a layer of width $l_{X} \, \simeq \, (n_{b}(H) \,
\sigma)^{-1} \,
\simeq \, 1.3 \times 10^{17} (T / 10^{6} K)^2
(\rho / 10^{-24} g \, cm^{-3})^{-1}
(T_{b} / 1000 K) \, cm$ by photons having $<h \nu> \, \sim \, 2.3 k_{B} T$.
This length scale defines a mass $M_{X} \, \simeq \, 2.1 \times 10^{-3}
(T / 10^{6} K)^{7} (\rho / 10^{-24} g \, cm^{-3})^{-2} (T_{b} / 1000 K)^{2} \,
M_{\odot}$. Therefore, blobs with $M_{blob} \, \leq \, M_{X}$ will be
heated throughout and the more massive ones only in a thin layer $\sim \,
l_{X}$
near their surfaces. This X-ray heating will kill the smallest blobs from
the original spectrum (eqs. 8 and 12), providing a possible explanation to the
lack of low-mass ($M \, \lappr \, 10^{-3} M_{\odot}$)
candidate events for MACHOs (Zylberajch 1994).
These figures must be considered as lower
limits to the heating rates since any other sources such as supernova
explosions or cosmic rays would increase $l_{X}$. However, we must stress
that the blobs are in principle able to
cool quickly to $ \lappr \, 100 K$
in a time-scale
$ \sim \, 3.3 \times 10^{3} \, [n_{b}(H) / n_b(H_{2})]^2 \, yr$
(for  $L_{H_{2}} \, \simeq \, 2 \times 10^{-29} \, {n_{b} (H_{2})}^2
\,  erg \, cm^{-3} \, s^{-1}$ for $n_{b}(H) \, \sim 10^{4} \, cm^{-3}$; see
Capuzzo-Dolcetta et al. 1995),
which is shorter than the protogalactic $t_{ff}$ at
that epoch. These low temperatures are enough for the blobs
 to become  gravitationally
unstable, and therefore it is not unlikely that a number of them can survive
until today.

Several other processes (not discussed here) may be also important for the
evolution of  the blobs (including ambipolar diffusion at
$ T_{b}  \, \leq \, 10^{4} K$
(e.g. Shu et al. 1987),
 and
$\sim \, 10^{8}$ collisions between blobs since their formation),
which may
bias the final mass distribution in a yet unknown manner.

The main result of these calculations is that
a thermally unstable regime of the protogalactic gas before
the standard stellar formation trigger induces the formation of
 condensations or blobs in a suitable mass interval.
If the physical conditions of the protogalactic gas are not very different
from the standard models of the protogalaxy, the most abundant surviving blobs
have typical masses $M \, \gappr \, 10^{-1} \,
M_{\odot}$, consistent with the microlensing candidates reported by the
MACHO and EROS collaborations. In this scenario, the MACHOs may be
identified with cold clouds which started to collapse from condensed
blobs because of the thermal instability of the gas. It has also implications
for the generation of low-mass Pop III luminous stars, a subject that has
received some attention in the recent past (e.g. Silk 1983,
Bessel \& Norris 1984 ;
Beers et al. 1985 ; Gass et al. 1988).

After the completion of this work we had notice of a recent work
(De Paolis et al. 1994) which discusses the issue of thermal instabilities
forming $M \, \sim \, 10^{6} \, M_{\odot}$ hot clouds later fragmenting
into low-mass objects identified with the MACHOs. Two comments are in
order. First, as stated by the authors, clustering of the lensing objects
is expected and this prediction can be checked against accumulated
observed events. Second, the clustered MACHOs formed inside those massive
clouds should in any case add up to the fraction resulting from the previous
epoch as discussed here; thus we may be observing different dark populations
originated closely in time but by different physics.

\noindent
\acknowledgements

We are indebted with the referee A. Ferrara for his  fruitful comments and
suggestions.
Discussions with R.Opher, S. Viegas,  A.C.S. Fria\c ca, A. Dal Pino Jr.,
 and G.A. Medina-Tanco
about various subjects of this work are also acknowledged.
J.E.H. acknowledges the warm hospitality given to him during a visit to
DAPNIA-Saclay and interesting discussions about the MACHO events maintained
with the EROS group. This work was partially supported by CNPq and
FAPESP grants.


\newpage

\begin{thebibliography}{}

\bibitem[1993]{A93}
\reference{} Alcock,C. et al. 1993, \nat , 365, 621

\bibitem[1995]{A95}
\reference{} Alcock,C. et al. , 1995, \prl , 74, 2867

\bibitem[1993]{Au93}
\reference{} Aubourg,E. et al. , 1993, \nat ,365, 623

\bibitem[1990]{Be90}
\reference{} Begelman,M.C. , 1990, in
"The interstellar medium of galaxies", eds.
H.A.Thompson,Jr. \& J.M.Shull, Kluwer Academic Publishers, 287

\bibitem[1985]{Bee85}
\reference{} Beers,T.C., Preston,G.W. \& Shectman,S.A., 1985, \aj, 90, 2089

\bibitem[1984]{bes84}
\reference{} Bessel,M.S. \& Norris,J. , 1984, \apj ,285, 622

\bibitem[1992]{Ca95}
\reference{} Capuzzo-Dolcetta, R., Di Fazio, A., \& Palla, F. 1995, in
"Physical Processes in Fragmentation and Star Formation", eds. R.
Capuzzo-Dolcetta et al., Reidel Publs.

\bibitem[1995]{Co95}
\reference{} Corbelli, E. \& Ferrara, A. 1995, \apj (July 10)

\bibitem[1972]{Da72}
\reference{} Dalgarno, A. \& McCray, R.A. 1972, ARA$\&$A, 10, 375

\bibitem[1995]{Dep95}
\reference{} De Paolis,F. et al., 1995, Comm. on Astrophys., in the press

\bibitem[1979]{Ei79}
\reference{} Eilek,J.A. \& Caroff,L.J. 1979, \apj , 233, 463

\bibitem[1979]{Fa85}
\reference{} Fall, S.M. \& Rees, M.J. 1985, \apj , 298, 18

\bibitem[1965]{Fi65}
\reference{} Field, G., 1965, \apj , 142, 531

\bibitem[1988]{G88}
\reference{} Gass,H., Liebert,J. \& Wehrse,R., 1988, \aap , 189, 194

\bibitem[1988]{Ga95}
\reference{} Gates, E.I., Gyuk, G. \& Turner, M.S., 1995, \prl, 74, 3724

\bibitem[1989]{B89}
\reference{} Gouveia Dal Pino,E.M. \& Opher,R., 1989a, \apj , 342, 686

\bibitem[1989]{B89a}
\reference{} Gouveia Dal Pino, E.M. \& Opher,R., 1989b, \mnras , 240, 573

\bibitem[1990]{B90}
\reference{} Gouveia Dal Pino,E.M. \& Opher,R., 1990, \aap , 231, 551

\bibitem[1991]{B91}
\reference{} Gouveia Dal Pino, E.M. \& Opher,R., 1991, \aap , 242, 319

\bibitem[1993]{B93}
\reference{} Gouveia Dal Pino, E.M. \& Opher,R., 1993, \mnras , 263, 687

\bibitem[1983]{Le83}
\reference{} Lepp, S. \& Shull, J.M. 1983, \apj, 270, 578

\bibitem[1993]{Mu93}
\reference{} Murray, S.D. et al. 1993, \apj, 407, 588

\bibitem[1995]{P95}
\reference{} Paczy\'nski, B. 1995, astro-ph/9508006

\bibitem[1977]{S94}
\reference{} Sahu, K.  1994, Nature, 370, 275

\bibitem[1977]{R77}
\reference{} Rees,M.J. \& Ostriker,J.P. 1977, \mnras , 179, 541

\bibitem[1987]{Shu87}
\reference{} Shu,F., Adams,F. e Lizano,S., 1987, \araa , 25, 23

\bibitem[1983]{S83}
\reference{} Silk,J., 1983, \mnras, 205, 705

\bibitem[1994]{U94}
\reference{} Udalski,A. et al. , 1994, Acta Astronomica 44, 165

\bibitem[1995]{V95}
\reference{}  Vietri,M. \& Pesce,E. 1995 (preprint)

\bibitem[1994]{Z94}
\reference{} Zylberajch, S., in Proceedings of the
the Vulcano Workshop, 1995, Mannocchi, G. \& Giovanelli, F.,Eds.,
Ed. Compositori, Bologna, 105

\end{thebibliography}
\end{document}